\documentclass[twocolumn, superscriptaddress, prb, showpacs]{revtex4}

\usepackage{graphicx}
\usepackage{amsmath}

\def\A{{\bf A}}
\def\B{{\bf B}}
\def\Be{{\bf B}^{\text{ext}}}

\def\M{{\bf M}}
\def\m{{\bf m}}
\def\p{{\bf p}}
\def\r{{\bf r}}

\begin{document}

\title{NMR shifts for polycyclic aromatic hydrocarbons from
first-principles}
\author{T. Thonhauser}
\affiliation{Department of Physics, Wake Forest University,
Winston-Salem, North Carolina 27109, USA.}
\author{Davide Ceresoli}
\affiliation{Department of Materials Science and Engineering, MIT,
Cambridge, Massachusetts 02139, USA.}
\author{Nicola Marzari}
\affiliation{Department of Materials Science and Engineering, MIT,
Cambridge, Massachusetts 02139, USA.}
\date{\today}

\begin{abstract}  
We present first-principles, density-functional theory calculations of
the NMR chemical  shifts for polycyclic aromatic hydrocarbons, 
starting with benzene and increasing sizes up to the one- and
two-dimensional infinite limits of graphene ribbons and sheets. Our
calculations are performed using a combination of the recently
developed theory of orbital magnetization in solids, and a novel
approach to NMR calculations where chemical shifts are obtained from
the derivative of the orbital magnetization with respect to a
microscopic, localized magnetic  dipole.  Using these methods we study
on equal footing the $^1$H and $^{13}$C shifts  in benzene, pyrene,
coronene, in naphthalene, anthracene, naphthacene, and pentacene, and
finally in graphene, graphite, and an infinite graphene ribbon. Our
results show very good agreement with experiments and allow us to
characterize the trends for the chemical shifts as a function of system
size.
\end{abstract}


\maketitle

\section{Introduction}
\label{sec:introduction}

The experimental technique of nuclear magnetic resonance (NMR) has been
known since the 1930's,\cite{Rabi} and has since become one of the most
widely used methods in structural chemistry.\cite{NMR_encyclopedia}
Along with experimental advancements over the several past decades, it
was soon realized that ab-initio calculations could greatly aid in
unambiguously determining structures from NMR spectra. Such
calculations were first developed in the quantum-chemistry
community;\cite{Kutzelnigg_90} however, these developments applied only
to finite systems, such as molecules and clusters. Application to
extended systems is hindered by the difficulty of including macroscopic
magnetic fields, which require a non-periodic vector potential and
therefore destroy Bloch symmetry. This shortcoming was overcome by
Mauri \emph{et al.}, who developed a linear-response approach for
calculating NMR shieldings in periodic crystals based on the\
long-wavelength limit of a periodic modulation of the applied magnetic
field.\cite{Mauri_96} Similarly, Sebastiani and Parrinello used the
Wannier representation to derive an alternative linear-response
approach based on the application of an infinitesimal uniform magnetic
field.\cite{Sebastiani_01} In recent years these approaches have been
advanced and refined,\cite{Pickard_Mauri_01,Pickard_Mauri_03,Yates_07}
leading to a growing use in modern plane-wave pseudopotential
codes.\cite{paratec,para2, castep} Nevertheless, the linear-response
framework central to all these approaches makes them fairly complex and
difficult to implement.

We have recently developed a novel method to calculate chemical shifts
in periodic systems,\cite{Thonhauser_07} where the need for a
linear-response framework is circumvented by construction. While early
results for simple molecules and hydrocarbons provided a proof of
principle for our approach,\cite{Thonhauser_07} albeit limited to
hydrogen shifts,  we apply it here systematically to calculate chemical
shifts for both hydrogen and carbon in polycyclic aromatic hydrocarbons
(PAH). Starting from benzene, we will explore the one-dimensional and
two-dimensional progression that include either pyrene and coronene on
one side, or naphthalene, anthracene, etc., until we reach the infinite
limits represented by graphene (and, by extension, graphite), or
graphene ribbons.

We focus this study on PAHs because of the recent surge of interest in
energy related materials and their impact on environmental and health
issues. PAHs occur in fossil fuels such as oil and coal, and they are a
byproduct of incomplete combustion in e.g. wood, fat, tobacco, or
incense. Depending on structure, these pollutants can be extremely
toxic, carcinogenic, mutagenic, and teratogenic. It is thus not
surprising that in recent years NMR techniques have been developed to
determine the content of PAHs in our environment.\cite{Weisshoff_02} A
tool combining ab-initio information with experimental techniques might
lead to more efficient and precise devices for detecting PAHs in e.g.\
soil or air.

This paper is organized in the following way: For completeness we
present in Sec.~\ref{sec:theory} a short theory summary of our converse
approach to NMR shifts. Details about our numerical calculations and
about the practical implementation of the converse method can be found
in Sec.~\ref{sec:details}. Results for finite PAHs are presented in
Sec.~\ref{sec:finite}, whereas results for periodic systems are
collected in Sec.~\ref{sec:infinite}. We will conclude and summarize in
Sec.~\ref{sec:conclusions}.

\section{Theory}
\label{sec:theory}

The usual approach to calculating NMR shifts in periodic systems is to
apply a magnetic field and calculate the local field at the nucleus
using a linear-response framework. We will refer to this approach as
\emph{direct}, since it computes the shifts directly from the applied
and induced fields. Since a constant field is not compatible with
periodic-boundary conditions, the approach developed by the solid-state
community has been to use linear-response theory in the limit of
long-wavelength perturbations.\cite{Mauri_96} We recently argued that
it is actually possible to calculate the shifts in periodic systems
without using a linear-response framework. In our \emph{converse}
approach we circumvent the linear-response framework in that we relate
the shifts to the macroscopic magnetization induced by magnetic point
dipoles placed at the nuclear sites of interest. We report here the
essential result and refer the reader to
Ref.~[\onlinecite{Thonhauser_07}] for details. 

We define  $E$ as the energy of a virtual magnetic dipole $\m_s$ at one
nuclear center $\r_s$ in the field $\B$ for a finite system, or as the
energy per cell of a periodic lattice of such dipoles.  Then, writing
the macroscopic magnetization as $M_\beta=-\Omega^{-1}\,\partial
E/\partial B_\beta$ (where $\Omega$ is the cell volume),
\begin{equation}\label{equ:converse}
\delta_{\alpha\beta} - \sigma_{s,\alpha\beta} =
 - \frac{\partial}{\partial B_\beta}
 \frac{\partial E}{\partial m_{s,\alpha}} =
 - \frac{\partial}{\partial m_{s,\alpha}}
 \frac{\partial E}{\partial B_\beta} =
 \Omega\frac{\partial M_\beta}{\partial m_{s,\alpha}}\;.
\end{equation}
It follows that $\tensor{\sigma}_s$ accounts for the shielding
contribution to the macroscopic magnetization induced by a magnetic
point dipole $\m_s$ sitting at nucleus $\r_s$ and all of its periodic
replicas. Instead of applying a constant (or long-wavelength) field
$\Be$ to an infinite periodic system and calculating the induced field
at all equivalent $s$ nuclei, we apply an infinite array of magnetic
dipoles to all equivalent sites $s$, and calculate the change in
magnetization. Since the perturbation is now periodic, it can simply be
computed using finite differences of ground-state calculations. Note
that $\M=\m_s/\Omega + \M^{\text{ind}}$, where the first term is
present merely because we have included magnetic dipoles by hand. The
shielding is related to the true induced magnetization via
$\sigma_{s,\alpha\beta}= -\Omega\,\partial
M^{\text{ind}}_\beta/\partial m_{s,\alpha}$.

NMR is a technique that probes for properties of the electronic
structure near the nuclei. Thus, a good description of the wave
functions near the nuclei is vital. The formalism described above can
directly be implemented into all-electron programs such as LAPW codes.
However, many codes use plane waves as a basis and the ionic potential
is replaced by a pseudopotential to keep the computational cost low. In
these cases, extra care has to be taken when calculating NMR shifts. 
While hydrogen shifts can be described correctly using Coulombic
(pseudo)potentials, a reconstruction of the pseudovalence wave functions
in the core region has to be performed for elements beyond the first
row.\cite{Mauri_96} This can be achieved with the so called
gauge-including projector augmented-wave (GIPAW) method, which has been
developed by Pickard and Mauri in 2001 and proved very efficient for
NMR calculations using pseudopotentials.\cite{Pickard_Mauri_03} We have
derived the corresponding GIPAW formalism for the converse method, but
the lengthy, mathematical details will be presented in a forthcoming
article.\cite{GIPAW}

\section{Implementation and Calculational details}
\label{sec:details}

We have implemented the converse method outlined above into the
plane-wave density functional theory code PWSCF, which is part of the
{\sc Quantum-ESPRESSO} distribution.\cite{pwscf} We added an extra term
to the Hamiltonian taking into account the electron orbital interaction
with a nuclear magnetic dipole $\m_s$ sitting at nucleus $\r_s$, by the
usual substitution for the momentum operator 
$\p\rightarrow\p+\frac{e}{c}\A$, where $\A$ is the vector potential of
a periodic array of nuclear dipoles. This is done conveniently in
reciprocal space and requires only a few dozen lines of additional
code. The calculation and implementation of the orbital
magnetization\cite{Resta_05,Thonhauser_05, Ceresoli_06,
Niu,Ceresoli_07} necessary to evaluate Eq.~(\ref{equ:converse}) is
somewhat more involved. However, a growing number of codes has the
calculation of the orbital magnetization already implemented so that
the NMR shifts can be calculated readily. Also, if one is only
interested in finite systems, one can calculate the induced magnetic
moment instead of the orbital magnetization, which is much simpler (the
moment can be calculated from the quantum-mechanical current, which
requires only knowledge of the wave functions). Since we are interested
in shifts of atoms other than hydrogen, we also implemented a GIPAW
reconstruction, as mentioned above.

It might appear that the converse method is computationally demanding,
since we need to perform $3N$ calculations to obtain the shielding
tensor for $N$ atoms. However, note that in practice a single full
self-consistent ground-state calculation is performed once, and then
the dipole perturbations according to each shielding are applied to
this ground state. Convergence after the perturbation is much faster
than the ground-state calculation, and we find that NMR shifts for
systems with hundreds of atoms can be easily calculated.

For our calculations in the following sections we have used a PBE
exchange-correlation functional and norm-conserving pseudopotentials of
the Troullier-Martins type\cite{TM} with a cutoff of 80 Rydberg. The
structural parameters of all systems have been optimized within this
framework. For the dipole perturbation we chose a value of $|\m_s|$ of
1$\mu_\mathrm{B}$, although the results are independent of the exact
value over a wide range.

In order to compare the accuracy of our converse method to the direct
method, we performed test calculations on benzene and diamond. For the finite
system benzene using the converse method we find an absolute hydrogen
shielding of 22.97~ppm and carbon shielding of $-165.2$~ppm. These
results compare well with the
results obtained using the direct method: 22.69~ppm and $-163.4$~ppm.
For periodic diamond we find a shielding of $-63.89$~ppm
compared to the direct method which gives
$-65.85$~ppm.\cite{Pickard_Mauri_01} Note that in all cases the carbon
shielding includes only the shielding from the valence electrons and
does not include the core shielding, which we calculate to be
+200.3~ppm. The small deviations between the converse and the direct
method are most likely due to the fact that the direct method uses a
linear-response framework in which a finite q-vector is used to make
the magnetic field periodic and modulate it over space, instead of
using a truly constant magnetic field.

\section{Results}
\label{sec:results}

NMR experiments usually report the isotropic shielding
$\sigma_s=\frac{1}{3}{\text{Tr}}[\,\tensor{\sigma}_s\,]$ via the
chemical shift $\delta_s=-(\sigma_s - \sigma_{\text{ref}})$. Here
$\sigma_{\text{ref}}$ is the isotropic shielding of a reference
compound such as tetramethylsilane (TMS). Note that we have not
calculated the shielding of TMS itself, however, we can still  report
shifts with respect to TMS by using our calculations for e.g.\ benzene
as an intermediate reference:
$\delta_s=-(\sigma_s-\sigma^{\text{benzene}}_{\text{calc}} -
\delta_{\text{TMS}}^{\text{benzene}})$. For 
$\delta_{\text{TMS}}^{\text{benzene}}$ we have used the experimental
values of 128.6~ppm for carbon and 7.26~ppm for hydrogen.\cite{landolt}
Henceforth, we will use the notation $\delta^{\text{molecule}}_s$ to
describe the shift relative to TMS of atom $s$ in a certain molecule.
We will use letters $s=A,B,C,\dots$ to refer to hydrogen shifts and
numerals $s=1,2,3,\dots$ for carbon shifts.

\begin{figure}
\begin{center}
\includegraphics[width=1.75cm]{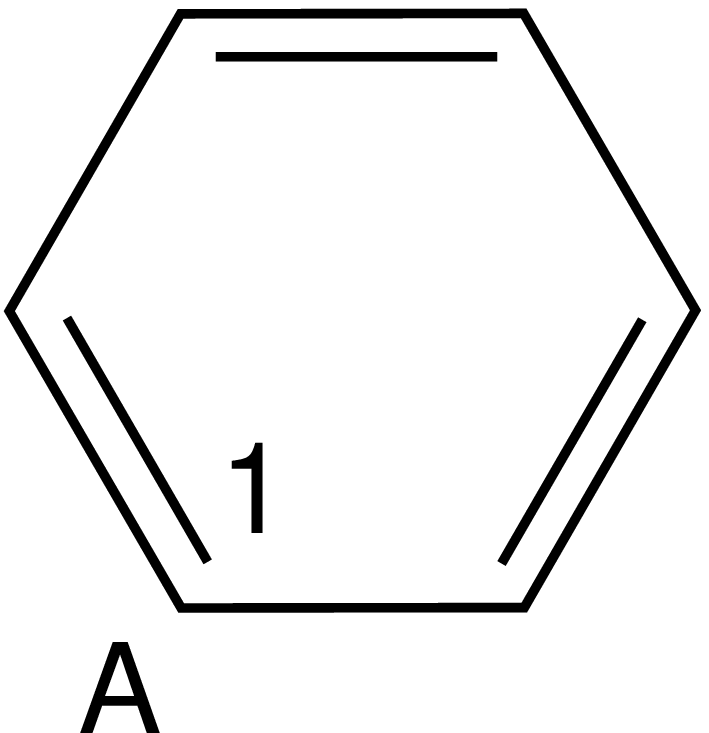}\hspace{1cm}
\includegraphics[width=5cm]{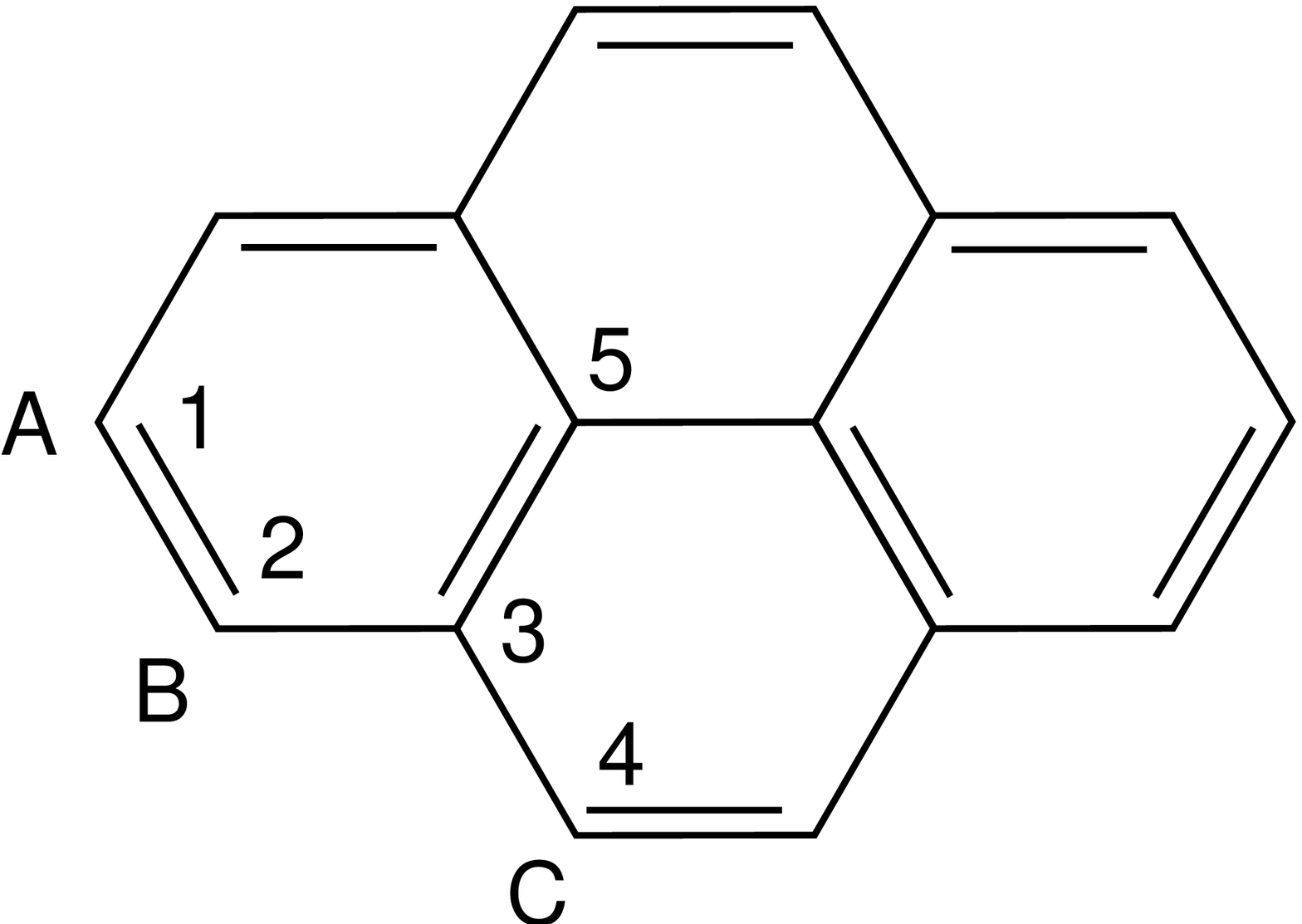}
\end{center}
\caption{\label{fig:pyrene} Schematic model of benzene (left) and pyrene
(right). In order to refer to certain atoms and their NMR shift, we use
the labeling shown. In general, hydrogen atoms are label by letters, whereas carbon atoms are labeled
by numerals.}
\end{figure}

\subsection{Finite systems}
\label{sec:finite}

We start out by calculating the $^1$H and $^{13}$C NMR shifts of
several small PAHs. We start with benzene, and move towards the
one-dimensional limit by adding one ring at a time, to study
naphthalene, anthracene, naphthacene,  and pentacene. The
two-dimensional limit is considered via pyrene and coronene, i.e.\
surrounding benzene by more rings in the plane. For all these
calculations we used a  large supercell with at least 16 Bohr of vacuum
between periodic replicas. Note that the magnetic susceptibility of
these systems effectively vanishes and our computed shifts can be
directly compared to the experimental ones without any shape
correction.\cite{note1}

\begin{figure}
\begin{center}
\includegraphics[width=4.5cm]{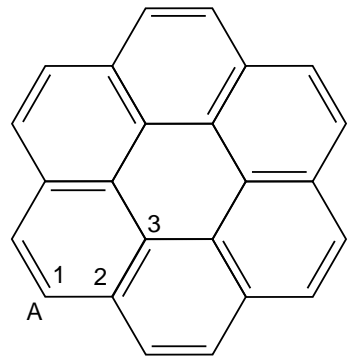}
\end{center}
\caption{\label{fig:coronene} Schematic model of coronene.}
\end{figure}

\begin{table}
\caption{\label{tab:finite_1} Hydrogen and carbon NMR chemical shifts 
in ppm for pyrene and coronene. For a definition of atomic positions
see Figs.~\ref{fig:pyrene} and \ref{fig:coronene}. Note that we have used
the shifts of benzene as a reference.}
\begin{tabular*}{\columnwidth}{@{}l@{\extracolsep{\fill}}ccr@{}}\hline\hline
Molecule&   atom        &  experiment      & calculation\\\hline
Benzene &   1           &  128.6$\;^a$     & ---        \\
        &   A           &   7.26$\;^a$     & ---        \\
Pyrene  &   1           &  125.5$\;^a$     & 122.2      \\
        &   2           &  124.6$\;^a$     & 122.1      \\
        &   3           &  130.9$\;^a$     & 130.4      \\
        &   4           &  127.0$\;^a$     & 126.0      \\
        &   5           &  124.6$\;^a$     & 123.4      \\
        &   A           &  7.60--8.15$\;^b$&  8.61      \\
        &   B           &  7.60--8.15$\;^b$&  8.57      \\
        &   C           &  7.60--8.15$\;^b$&  8.47      \\
Coronene&   1           &  126.2$\;^a$     & 123.5      \\
        &   2           &  128.7$\;^a$     & 129.3      \\
        &   3           &  122.6$\;^a$     & 122.6      \\
        &   A           &   8.89$\;^c$     &  9.78      \\\hline\hline
\end{tabular*}
\raggedright
$^a\;$Reference [\onlinecite{landolt}].\\
$^b\;$Reference [\onlinecite{sadtler}].\\
$^c\;$Reference [\onlinecite{Pouchert}].
\end{table}

Our results for pyrene and coronene are collected in
Table~\ref{tab:finite_1}. For a definition of their atomic positions
see Figs.~\ref{fig:pyrene} and \ref{fig:coronene}. In general, we find
very good agreement between the experimental results and our
calculations. The small deviations are most likely due to the
approximative nature of DFT with respect to the exchange and
correlation functional. Using e.g.\ the local-density approximation for
the exchange-correlation functional,  we get slightly different
results. However, a detailed and comprehensive study is necessary to
adequately investigate the effect of the functional used on the NMR
shifts.\cite{Zhao_08} In principle, since we are calculating the shift
through the orbital magnetization, which is closely related to
electrical currents, it might be more appropriate to use a
current-density functional rather than one of the standard density-only
functionals. Unfortunately, appropriate current-density functionals
(e.g.\ Ref.~[\onlinecite{Vignale_87}]) are not yet available in most
DFT codes. Note that the relative shifts (that is, the difference in
shift between different atoms) are closely reproduced. This is
important, as relative shifts play the key role in linking NMR data to
a structure.

As one would expect, all shifts of pyrene and coronene are
significantly different from the pure benzene shifts. However, it is
interesting to see that the hydrogen shifts increase, while all carbon
shifts decrease---except one. It is $\delta^{\text{pyrene}}_3$ and
$\delta^{\text{coronene}}_2$ that rise above the 128.6~ppm of benzene.
These shifts both correspond to carbon atoms that have no hydrogens
attached.

\begin{figure}
\begin{center}
\includegraphics[width=2.2cm]{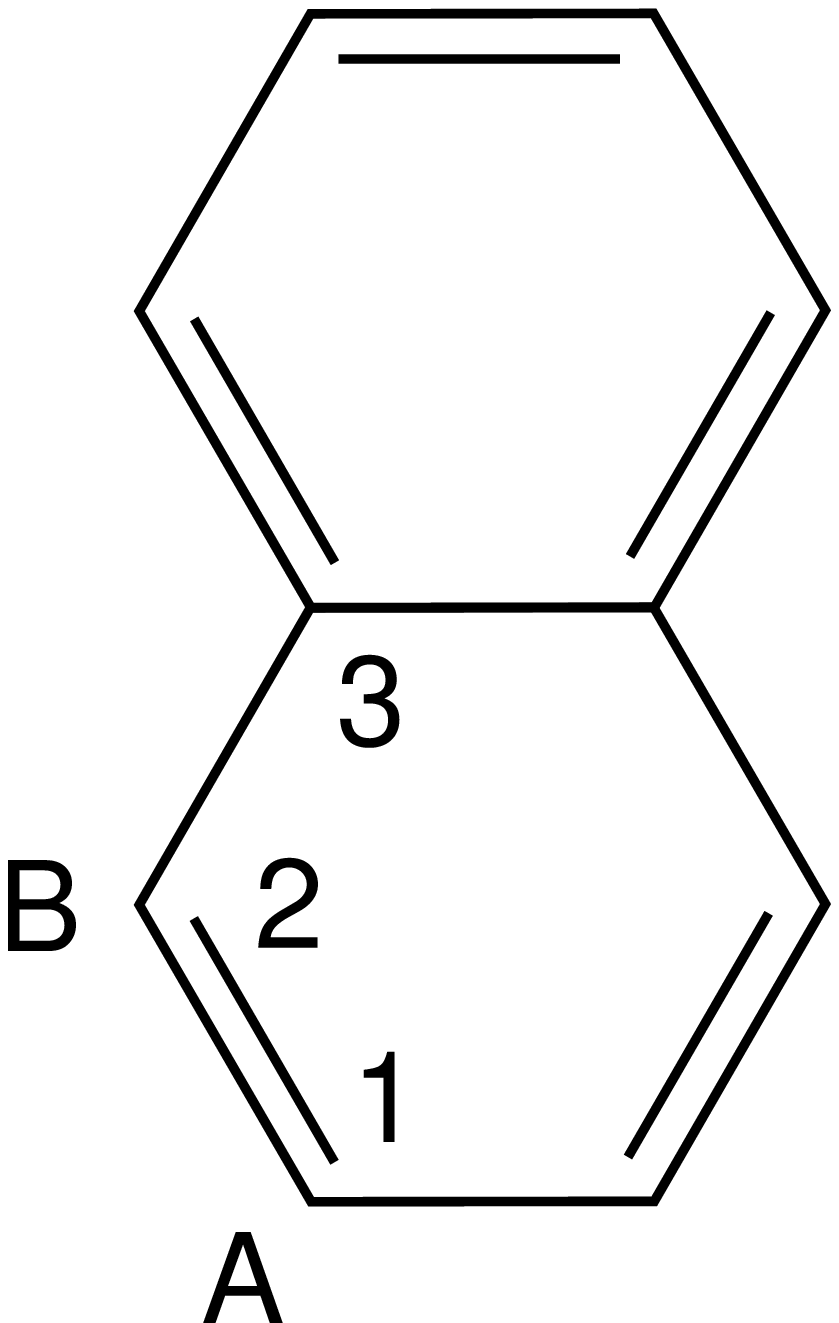}\hspace{1cm}
\includegraphics[width=2.2cm]{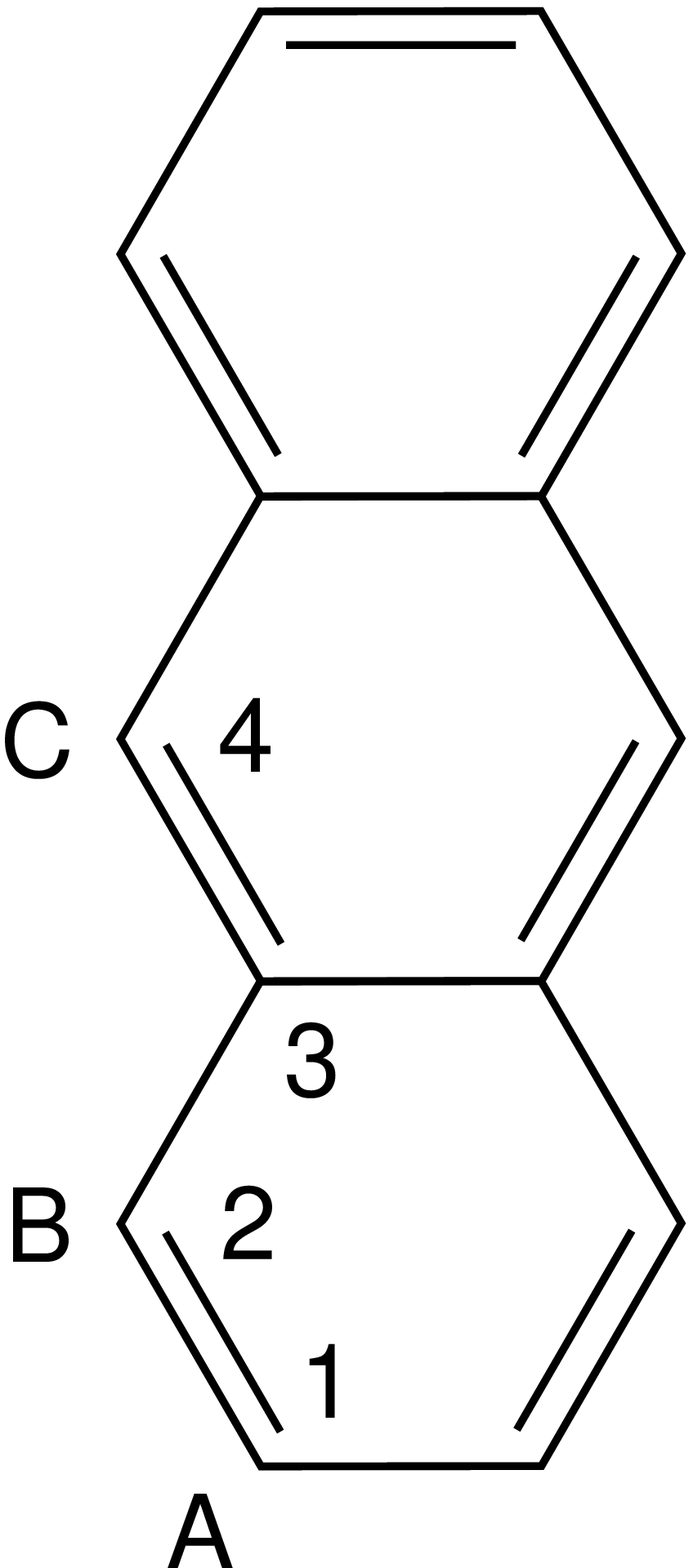}
\end{center}
\caption{\label{fig:naphthalene} ({\bf left}) Naphthalene. ({\bf right}) 
Anthracene.}
\end{figure}

\begin{table}
\caption{\label{tab:finite_2} Hydrogen and carbon NMR chemical shifts 
in ppm for naphthalene and anthracene. For a definition of atomic positions
see Fig.~\ref{fig:naphthalene}.}
\begin{tabular*}{\columnwidth}{@{}l@{\extracolsep{\fill}}ccr@{}}\hline\hline
Molecule  &   atom        &  experiment       & calculation\\\hline
Benzene   &   1           &   128.6$\;^a$     & ---        \\
          &   A           &    7.26$\;^a$     & ---        \\
Naphthalene&  1           &   125.7$\;^c$     & 123.0      \\
          &   2           &   127.8$\;^c$     & 126.2      \\
          &   3           &   133.4$\;^c$     & 133.1      \\
          &   A           &    7.41$\;^b$     & 8.08       \\
          &   B           &    7.79$\;^b$     & 8.13       \\
Anthracene&   1           &   126.1$\;^c$     & 125.0      \\
          &   2           &   128.1$\;^c$     & 127.8      \\
          &   3           &   131.6$\;^c$     & 132.0      \\
          &   4           &   125.3$\;^c$     & 121.6      \\
          &   A           &    7.41$\;^b$     & 7.95       \\
          &   B           &    7.95$\;^b$     & 8.24       \\
          &   C           &    8.40$\;^b$     & 8.44       \\\hline\hline
\end{tabular*}
\raggedright
$^a\;$Reference [\onlinecite{landolt}].\\
$^b\;$Reference [\onlinecite{sadtler}].\\
$^c\;$Reference [\onlinecite{Pouchert}].
\end{table}

\begin{figure}
\begin{center}
\includegraphics[width=2.2cm]{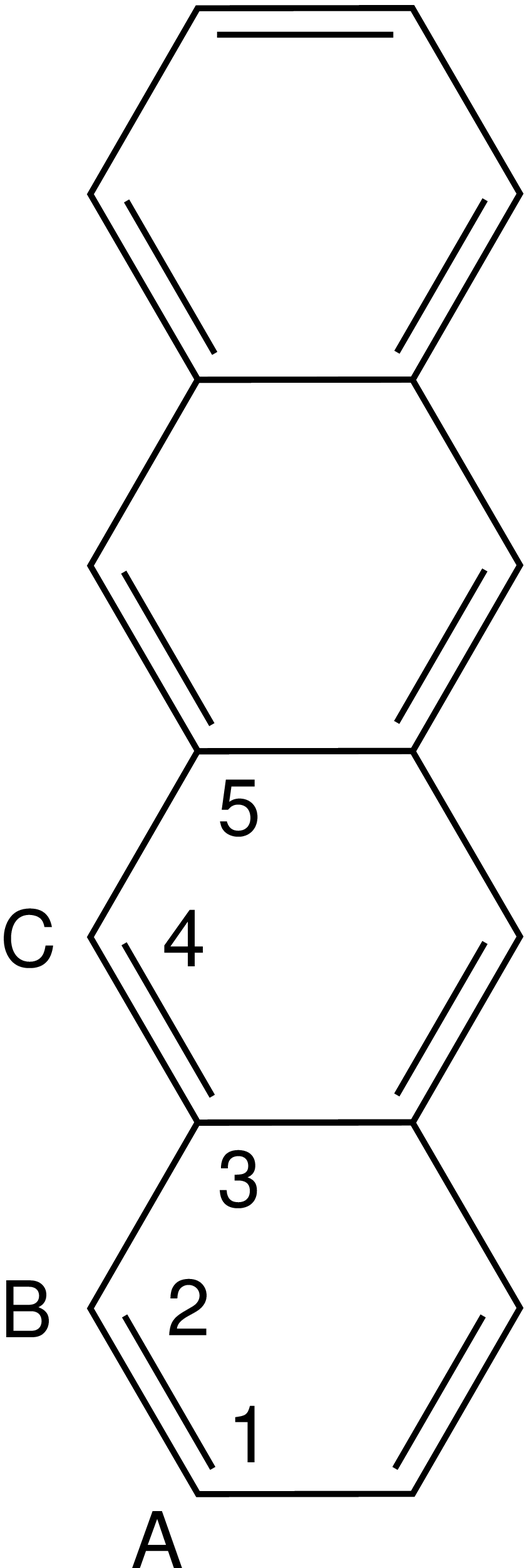}\hspace{1cm}
\includegraphics[width=2.2cm]{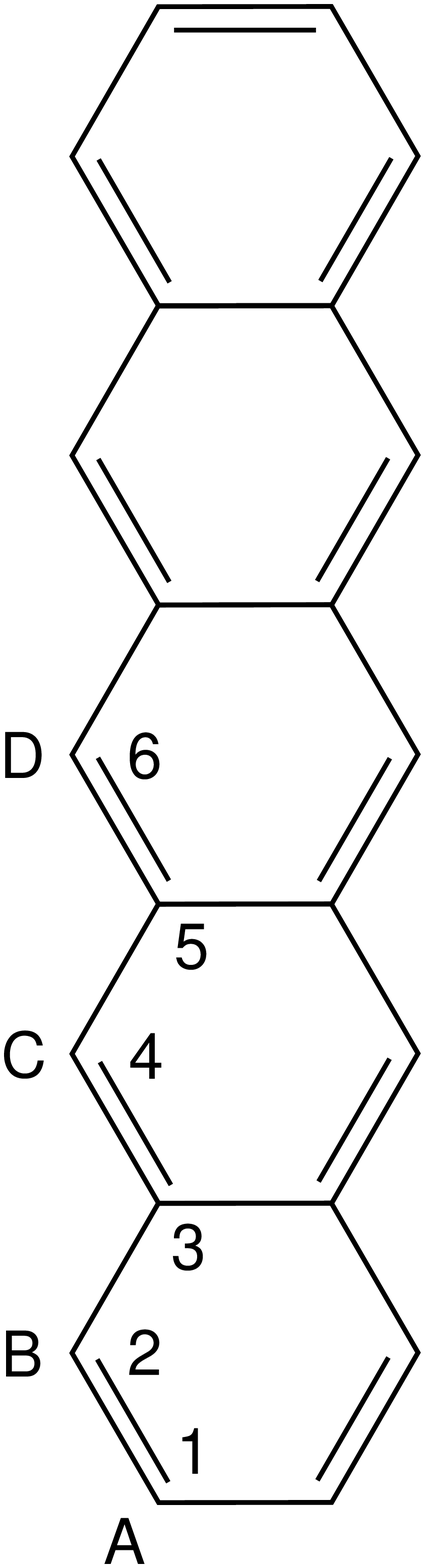}
\end{center}
\caption{\label{fig:naphthacene} ({\bf left}) Naphthacene. ({\bf right}) 
Pentacene.}
\end{figure}

Our results for the one-dimensional chains from naphthalene to
pentacene are given in Tables~\ref{tab:finite_2} and
\ref{tab:finite_3}. For a definition of the corresponding atomic
positions see Figs.~\ref{fig:naphthalene} and \ref{fig:naphthacene}. As
before, we find very good agreement between the experimental results
and our calculations. The exact same statements concerning the
differences compared to benzene apply: hydrogen shifts  increase, while
carbon shifts decrease---expect one. In these chains it is the carbon
shift of atom 3 that increases, which again corresponds to a carbon
that is not connected to a hydrogen.

Note that we have not included experimental results for naphthacene and
pentacene. Although there are some results reported (see e.g.\
Ref.~[\onlinecite{jpn}]) these molecules are insoluble in most of the
solvents used for the preparation of NMR cuvettes.

\subsection{Periodic systems}
\label{sec:infinite}

The systems discussed above can systematically be expanded until they
eventually become infinite. In a one-dimensional way, we extend the
chains naphthalene, anthracene, naphthacene, pentacene, \dots\ until we
reach an infinite ribbon. On the other hand, if we extend pyrene and
coronene in a two-dimensional way, we arrive at graphene. Stacking
graphene sheets on top of each other yields graphite.

\begin{table}
\caption{\label{tab:finite_3} Hydrogen and carbon NMR chemical shifts 
in ppm for naphthacene and pentacene. For a definition of atomic positions
see Fig.~\ref{fig:naphthacene}.}
\begin{tabular*}{\columnwidth}{@{}l@{\extracolsep{\fill}}ccr@{}}\hline\hline
\multicolumn{2}{c}{Naphthacene} & \multicolumn{2}{c}{Pentacene}\\
atom &  calc.  &  atom & calc.    \\\hline
1    &  122.3  &   1   & 123.3    \\
2    &  126.5  &   2   & 127.4    \\
3    &  130.5  &   3   & 130.6    \\
4    &  125.5  &   4   & 125.0    \\
5    &  128.1  &   5   & 128.5    \\
A    &  8.31   &   6   & 125.4    \\
B    &  8.68   &   A   & 8.37     \\
C    &  9.07   &   B   & 8.74     \\
     &         &   C   & 9.21     \\
     &         &   D   & 9.46     \\\hline\hline
\end{tabular*}
\end{table}

We have calculated the hydrogen and carbon shifts for these extended
systems. These infinite systems  show metallic behavior and therefore
require a dense k-point mesh.  For our calculations we have used up to
16 k-points in the directions of periodicity and a gaussian smearing of
0.001~Rydberg to obtain converged results.  As another result of the
metallic behavior, Knight shifts might influence the NMR spectrum.
While schemes do exist to calculate Knight shifts within plane-wave DFT
approaches,\cite{Nicola} we did not include them in our calculations.

\begin{figure}
\begin{center}
\includegraphics[width=2.2cm]{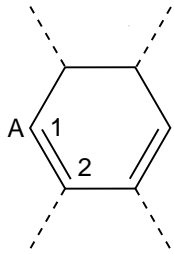}
\end{center}
\caption{\label{fig:ribbon} Section of an infinite ribbon consisting of
benzene rings.}
\end{figure}

For graphene we find a carbon shift of 118.0~ppm. The two in-equivalent
carbon shifts in graphite give 124.3~ppm and 134.9~ppm. For the
infinite ribbon depicted in Fig.~\ref{fig:ribbon} we find
$\delta^{\text{ribbon}}_A$=8.56~ppm,
$\delta^{\text{ribbon}}_1$=128.0~ppm, and
$\delta^{\text{ribbon}}_2$=132.2~ppm. Experimentally, solid state NMR
spectra of graphite show a broad peak in the range 155$\div$179~ppm,
depending on sample preparation. This broadening is due to a fairly
long lattice relaxation time $T_1$ and the presence of conduction
electrons.\cite{nmr_graphite}

In order to compare our results for graphite with experimental data, in
principle one would have to include a shape correction involving the
susceptibility.~\cite{note1} This correction is estimated to be of the
order of 1$\div$2~ppm in most forms of graphite sample such as powder
samples. The only exception is highly oriented pyrolytic graphite,
which has a very large and anisotropic magnetic susceptibility, where
the shape correction is expected to be of the order of 10$\div$20~ppm.

The calculated shifts for these periodic systems are interesting in and
of themselves. However, we now want to focus on the change in shift as
the systems grow larger and eventually become infinite.   One would
expect to find a correlation between the shifts in finite and infinite
systems. Indeed, looking at the carbon shifts we find
$\delta_{1}^{\text{benzene}}\to\delta_{5}^{\text{pyrene}}\to\delta_{3}^{\text{coronene}}\to\delta^{\text{graphene}}$
with the results $128.6 \to 123.4 \to 122.6 \to 118.0$~ppm. Although
NMR is a local probe, it turns out that the size of coronene is not yet
big enough for its bulk atoms to behave like graphene.

If we start at benzene again and add rings in a liner fashion, we can
grow a chain of arbitrary length. Looking at the carbon shifts we find
the sequence 
$\delta_{4}^{\text{anthracene}}\to\delta_{6}^{\text{pentacene}}\to\delta_{1}^{\text{ribbon}}$ 
with the corresponding results $121.6\to 125.4\to 128.0$~ppm. As with
the graphene sheet, this result suggests that a finite chain would need
to be longer than pentacene so that its carbon atoms behave like an
infinite ribbon. Looking at sequences that end at
$\delta^{\text{ribbon}}_A$ or $\delta^{\text{ribbon}}_2$, no clear
trend is visible.

\section{Conclusions}
\label{sec:conclusions}

We have used an alternative first-principles method to calculate NMR
chemical shifts of a variety of polycyclic aromatic hydrocarbons and
related infinite systems. Our results are in good agreement with
experiment. By going from finite to periodic systems, we observe trends
in shifts which suggest that neither coronene nor pentacene are large
enough to model their infinite counterparts.

In future work we would like to include the shifts of other PAHs as
well as study the Knight shifts of graphene.

\section{Acknowledgments}
\label{sec:acknowledgments}

We are grateful for extensive discussions with F.~Mauri. This work was
supported by NSF grant DMR-0549198, ONR grant N00014-07-1-1095, and the
DOE/SciDAC project on Quantum Simulation of Materials and
Nanostructures.



\begin{thebibliography}{99}
\bibitem{note1} It can be shown that the shielding
   for a general shape is related to our calculated one by
   $\sigma_{s,\alpha\beta}^{\text{shape}}\simeq\sigma_{s,\alpha\beta} -
   \delta_{\alpha\beta}\,4\pi\chi(1-n_\beta)$, where ${n}_\alpha$ are
   the depolarization coefficients (with $\sum_\alpha {n}_\alpha=1$)
   and $\chi$ is the magnetic susceptibility $\chi$.
\end{thebibliography}
\end{document}